\documentclass{elsart}
\usepackage{natbib,epsfig}
\begin{document}
%\runauthor{Angela C. Taylor}

%%%%%%%%%%%%%%%%%%%%%%%%%%%%%%%%%%%%%%%%%%%%%%%%%%%%%%%%%%%%%%%%%%%%%%%%%%

\begin{frontmatter}
\title{Observing the CMB at High-$\ell$ using the VSA and AMI}
\author[Angela]{Angela C. Taylor}
\address[Angela]{Astrophysics Group, Cavendish Laboratory, Madingley
  Road, Cambridge, UK}
\ead{act21@cam.ac.uk}

\begin{abstract}
We discuss two experiments - the Very Small Array (VSA) and the
Arcminute MicroKelvin Imager (AMI) - and their prospects for observing the
CMB at high angular multipoles. Whilst the VSA is primarily designed
to observe primary anisotropies in the CMB, AMI is designed to image
secondary anisotropies via the Sunyaev-Zel'dovich effect. The combined
$\ell$-range of these two instruments is between $\ell = 150$ and $\sim 10000$.
\end{abstract}

\end{frontmatter}

%%%%%%%%%%%%%%%%%%%%%%%%%%%%%%%%%%%%%%%%%%%%%%%%%%%%%%%%%%%%%%%%%%%%%%%%%%

\section{Introduction}

With the publication of the WMAP data \cite{wmap}, the focus for
observational CMB effort is moving to higher resolution. Here we
describe two experiments, one currently operating, one under
construction, which are designed to provide complementary observations
of primary and secondary CMB anisotropies over the multipole range
$\ell = 150$ to $\sim 10000$.

\section{The Very Small Array}

The Very Small Array (VSA) is a 14-element interferometric telescope
designed to image faint structures in the CMB on degree and sub-degree
angular scales. The instrument is situated on Mount Teide in Tenerife,
and is operated jointly by the Cavendish Astrophysics Group,
Cambridge, Jodrell Bank Observatory (JBO), Manchester and the Instituto de
Astrofisica (IAC), Tenerife. The VSA currently can operate in the
frequency range 26--36~GHz with an observing bandwidth of 1.5~GHz and
has been routinely observing the CMB since September 2000. 

\begin{figure}[t]
\centerline{
\psfig{figure=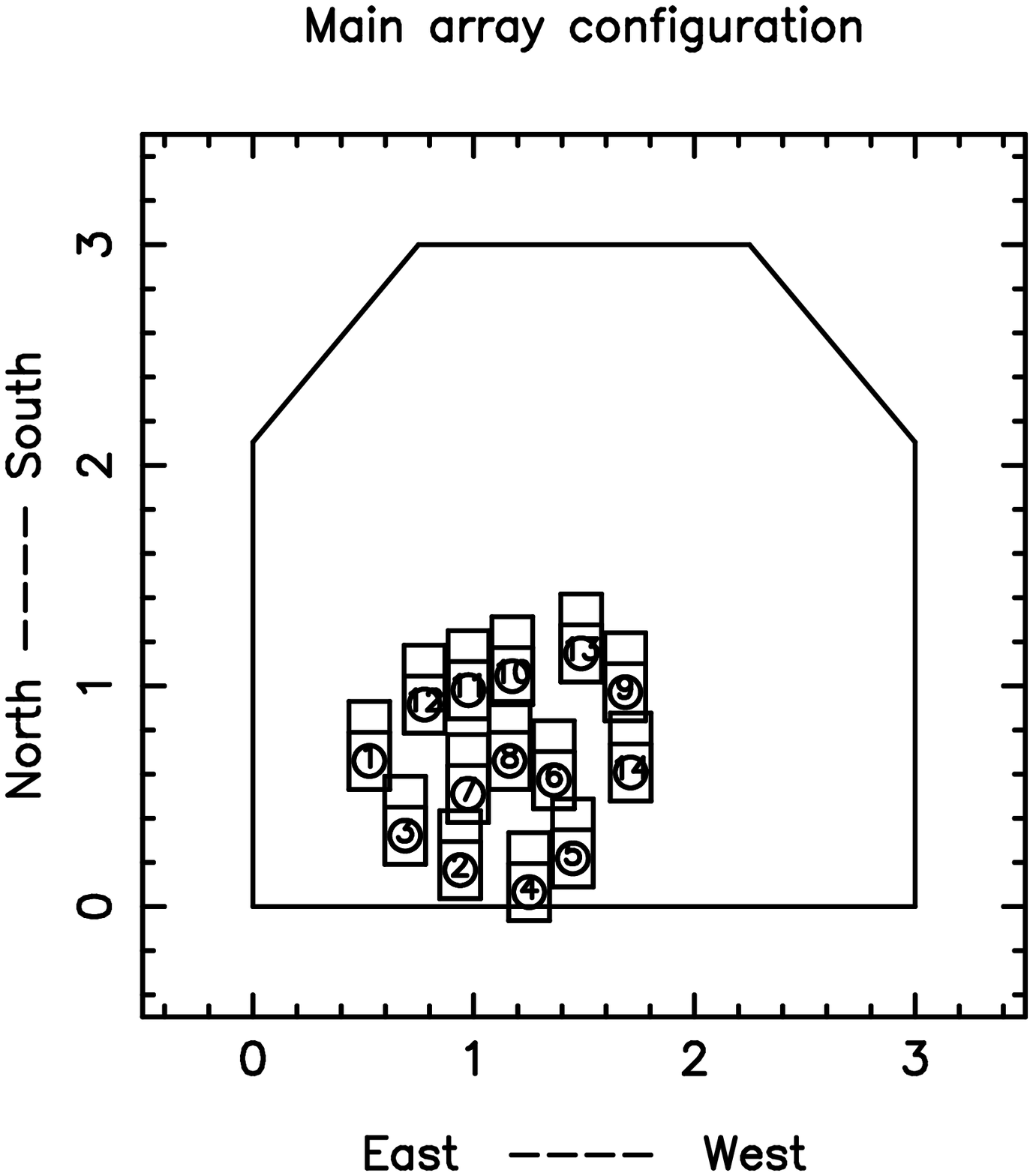,angle=-90,width=5cm,clip}
\quad
\psfig{figure=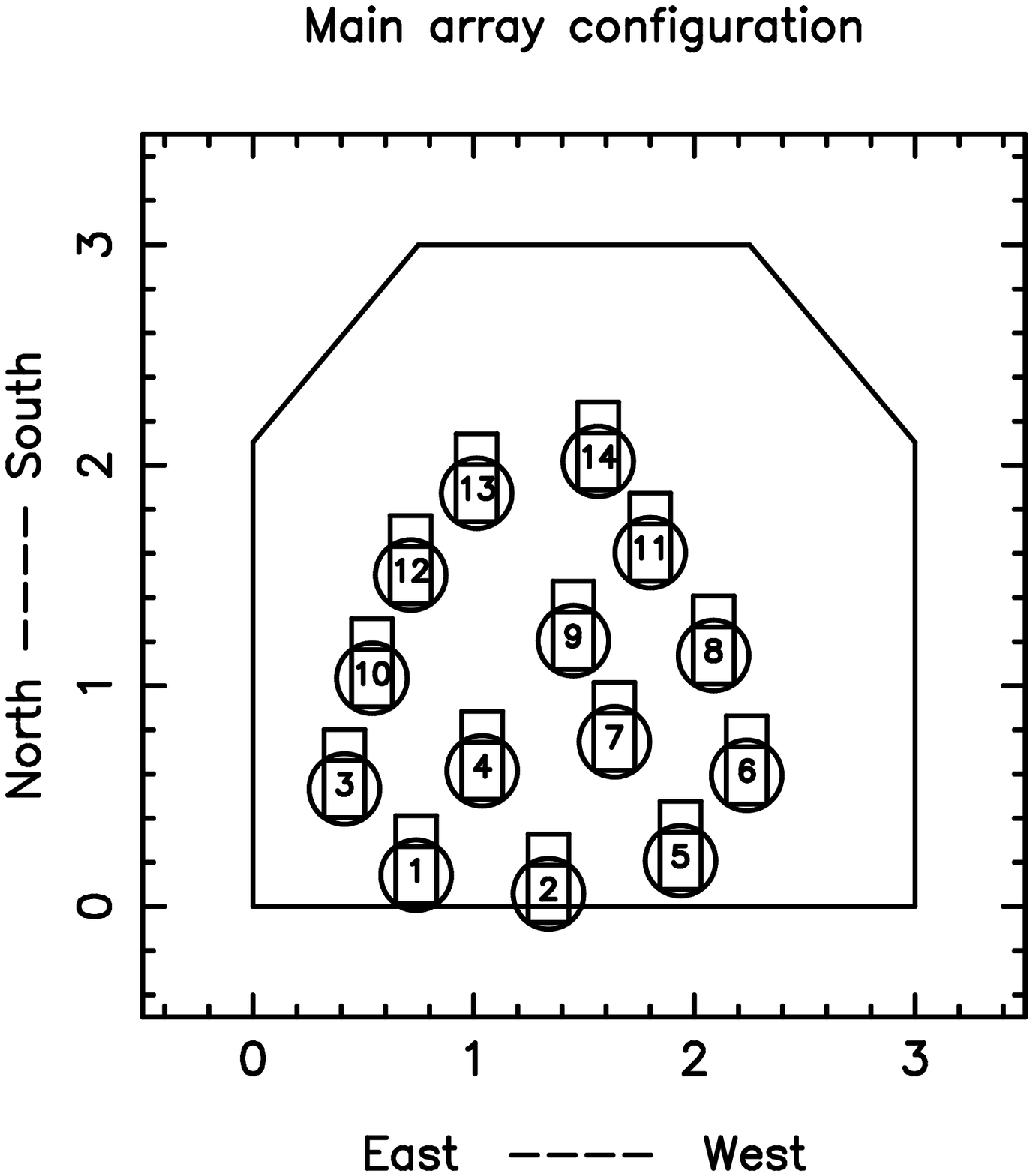,angle=-90,width=5cm,clip}
}
\caption{Left: Compact array configuration used during the first
  season of VSA observations. Right: extended array configuration,
  September 2001 - present.
\label{fig:configs}}
\end{figure}

Each receiving element of the array consists of a corrugated
horn-reflector antenna feeding a cooled HEMT amplifier.  The antennas
are mounted on a tilt-table and can be placed anywhere on the table,
allowing freedom to design the array configuration for specific
observational goals.  During the first two observing seasons
(September 2000 -- September 2001 and October 2001 -- present)
observations of the CMB have been made using two different array
configurations; the first used a compact configuration with a maximum
baseline (antenna separation) of 1.2~m, while during the second
season a more extended configuration was used, with a maximum baseline
of 2.5~m (see Figure \ref{fig:configs}).  The results from the two observing
seasons are summarised below.

\subsection{Observations in compact configuration}

\begin{figure}[t]
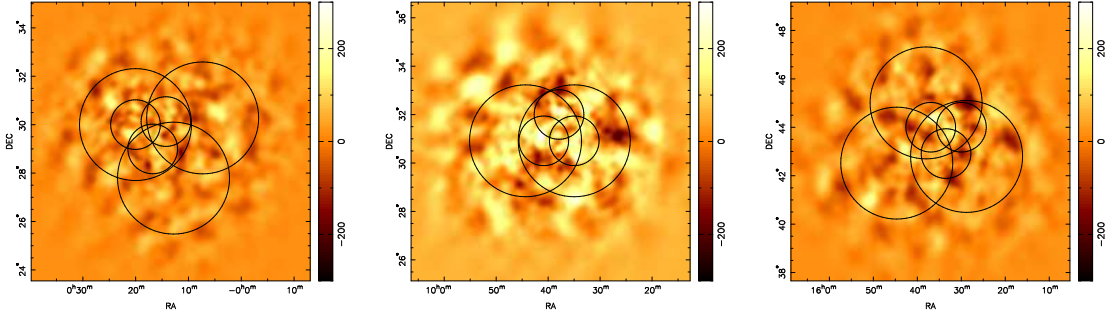

\centerline{
\psfig{figure=VSA1all.ps,angle=-90,width=4.5cm}
\quad
\psfig{figure=VSA2all.ps,angle=-90,width=4.5cm}
\quad
\psfig{figure=VSA3all.ps,angle=-90,width=4.5cm}
}
\caption{Source-subtracted maximum-entropy
reconstructions of each VSA mosaiced
region observed in both the compact and extended array configurations. 
Left: VSA1,1A and 1B. Middle: VSA2 and VSA2-OFF. Right: VSA3,3A and
3B. The black circles show the FWHM of the primary beam for each field
in the mosaic (FWHM $=4.6^\circ$ compact configuration, $2.04^\circ$
extended configuration). The maps are in units of $\mu$K. 
\label{fig:maps}}
\end{figure}

In its compact configuration, the VSA was sensitive to angular scales
of $3.6^\circ$--$0.4^\circ$ (equivalent to angular multipoles $\ell =
150 - 900$). It mapped CMB fluctuations at 34~GHz in eight fields
covering three separate areas of sky with a total area of 101 deg$^2$
(see Figure \ref{fig:maps}).  The noise levels in each field agree
well with the expected thermal noise level of the telescope, and there
is no evidence of any residual systematic features. Discrete radio
sources were detected using a separate 15~GHz survey \cite{9c} and
their effects removed using pointed follow-up observations at 34~GHz
using a single-baseline interferometer operating alongside and
simultaneously with the CMB observations. The residual confusion noise
due to unsubtracted radio sources is less than $15\,\mu$K in the
full-resolution images, which added in quadrature to the thermal noise
increases the noise level by 6 $\mu$K.  The {\em rms} contribution to
the images from diffuse Galactic emission is less than $6\, \mu$K.
The power spectrum derived from this data produced clear detections of
first and second acoustic peaks at $\ell{\approx 250}$ and
$\ell{\approx 550}$. The results of the first season of VSA
observations were released in May 2002, and presented in references
\cite{V1,V2,V3,V4}.

\subsection{Observations in extended configuration}

In September 2001, the VSA was re-configured for making
higher-resolution observations of the CMB. Each receiver was re-fitted
with a larger horn-reflector, 322~mm in diameter, and the antennas
were spaced more widely on the table.  In this `extended'
configuration, the VSA is sensitive to angular scales of
$1.8^\circ$--$0.25^\circ$ ($\ell = 300$--$1400$). By March 2002, it
had completed nine fields covering three separate regions of sky with
a total area of 25 deg$^2$ (these fields are contained within the same
regions of sky already observed by the VSA in its compact
configuration). 

\begin{figure}[t]
\begin{center}
\psfig{figure=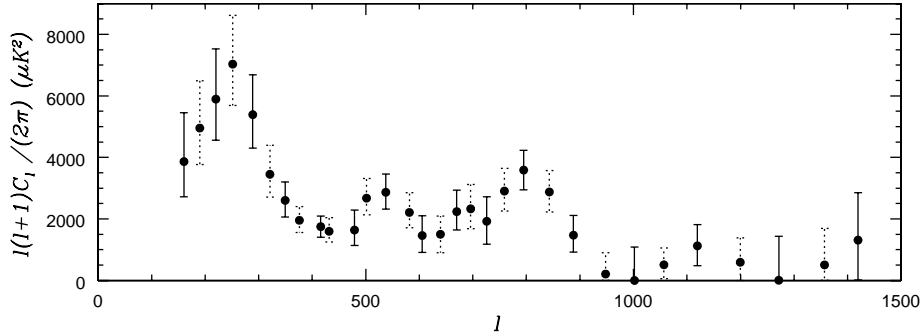,angle=-90,width=13cm}
\caption{Combined CMB power spectrum from compact and extended VSA
configurations. The error-bars represent $1\sigma$ limits; the two sets of
      data points correspond to alternative interleaved binnings of the data.
\label{fig:powerspectrum}}
\end{center}
\end{figure}

The combined power spectrum from the two observing seasons
(Figure~\ref{fig:powerspectrum}, \cite{V5}) covers angular multipoles
$\ell = 160$ -- $1400$, clearly resolves the first three acoustic
peaks, shows the expected fall off in power at high $\ell$ and starts
to constrain the position and height of a fourth peak.
Using the data from the two VSA configurations, we estimated
cosmological parameters for four different models of increasing
complexity \cite{V6}.  In each case, the Bayesian evidence was
calculated in order to determine whether the increased complexity of
the models was required by the data.  We found that the data are
adequately explained by a simple flat $\Lambda\rm{CDM}$ cosmology
without tensor modes. In this case, combining just the VSA and COBE
data sets yields the 68 per cent confidence intervals
$\Omega_{\rm{b}}h^2=0.034 ^{+0.007} _{-0.007}$,
$\Omega_{\rm{dm}}h^2=0.18 ^{+0.06}_{-0.04}$, $h=0.72^{+0.15}_{-0.13}
$, $n_{\rm s}=1.07 ^{+0.06 }_{-0.06}$ and $\sigma_8=1.17 ^{ +0.25 }_{
  -0.20}$. We also find that, by combining VSA data with estimates of
$H_0$ from either the HST Key Project \cite{hst} or Sunyaev-Zel'dovich
and X-ray observations \cite{jones, reese} we obtain strong evidence
for a non-zero cosmological constant independent of supernovae data.

The extended configuration results described above were announced in December
2002 and are presented in references \cite{V5,V6}. 

Observations in the extended configuration are ongoing, with
the aim of imaging a further 24 fields in 7 regions of the sky. This
programme is due to be completed by mid-2003, and will contain $\sim
4$ times the amount of data presented in \cite{V5}.
\subsection{Future VSA Observations}

The first three acoustic peaks in the CMB power spectrum have now been
detected at high significance, with positions and heights in line with
those predicted by the concordance model. However, significant
degeneracies in estimates of cosmological parameters still exist,
which can be broken by making higher resolution measurements, beyond
$\ell=900$. Recent results from the Cosmic Background Imager (CBI)
\cite{CBI} and Arcminute Cosmology Bolometer Array Receiver (ACBAR)
\cite{ACBAR} at intermediate and high $\ell$-values have shown the
gradual decline of the CMB power spectrum towards higher $\ell$, which
is expected from Silk damping, but beyond $\ell \approx 2000$,
the CBI data show an unexpected detection of power, which may be due
to the integrated SZ effect \cite{CBI-bond}.
  
\begin{figure}
\centerline{
\epsfig{file=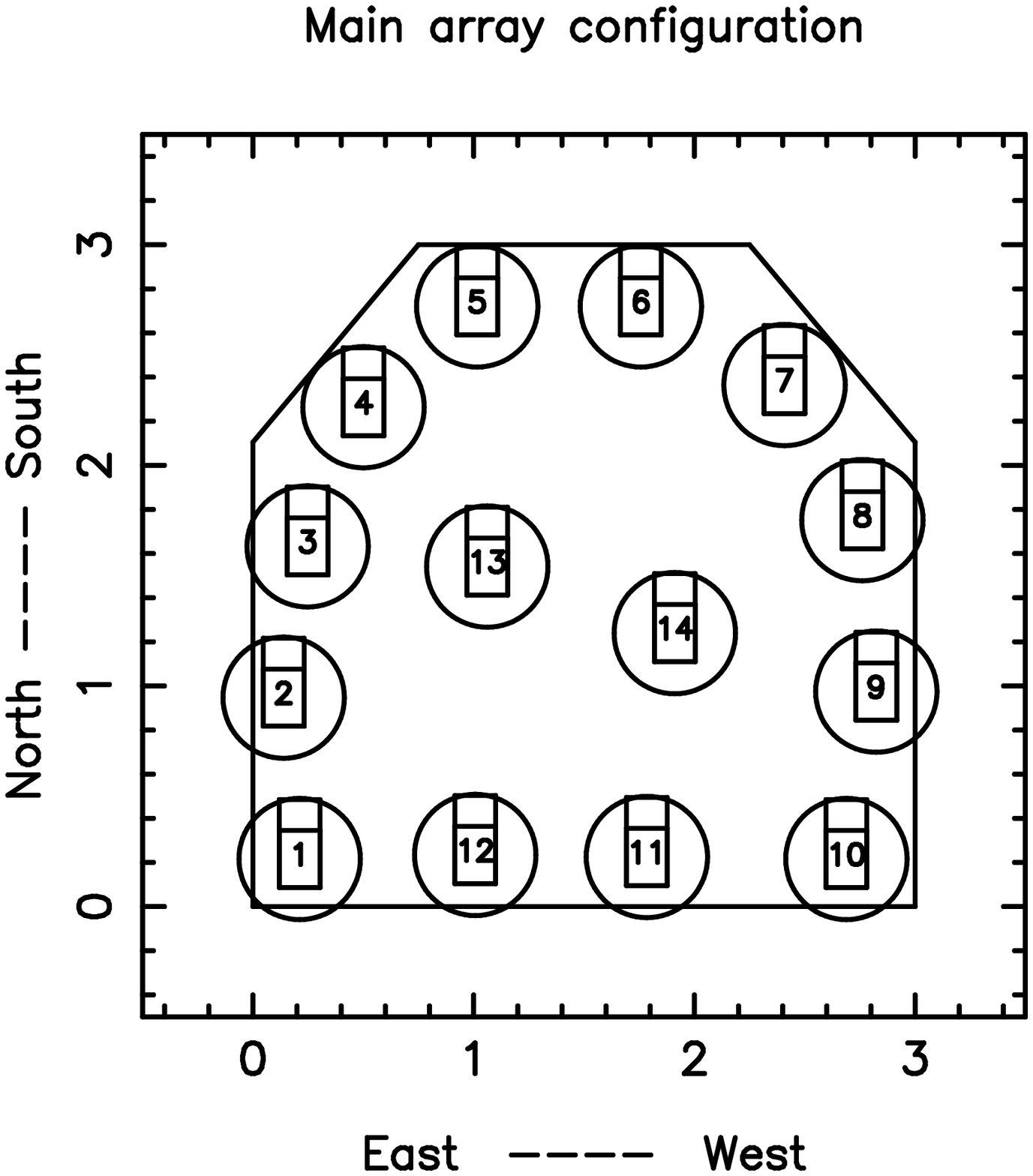,angle=-90,width=5cm,clip}
\quad
\epsfig{file=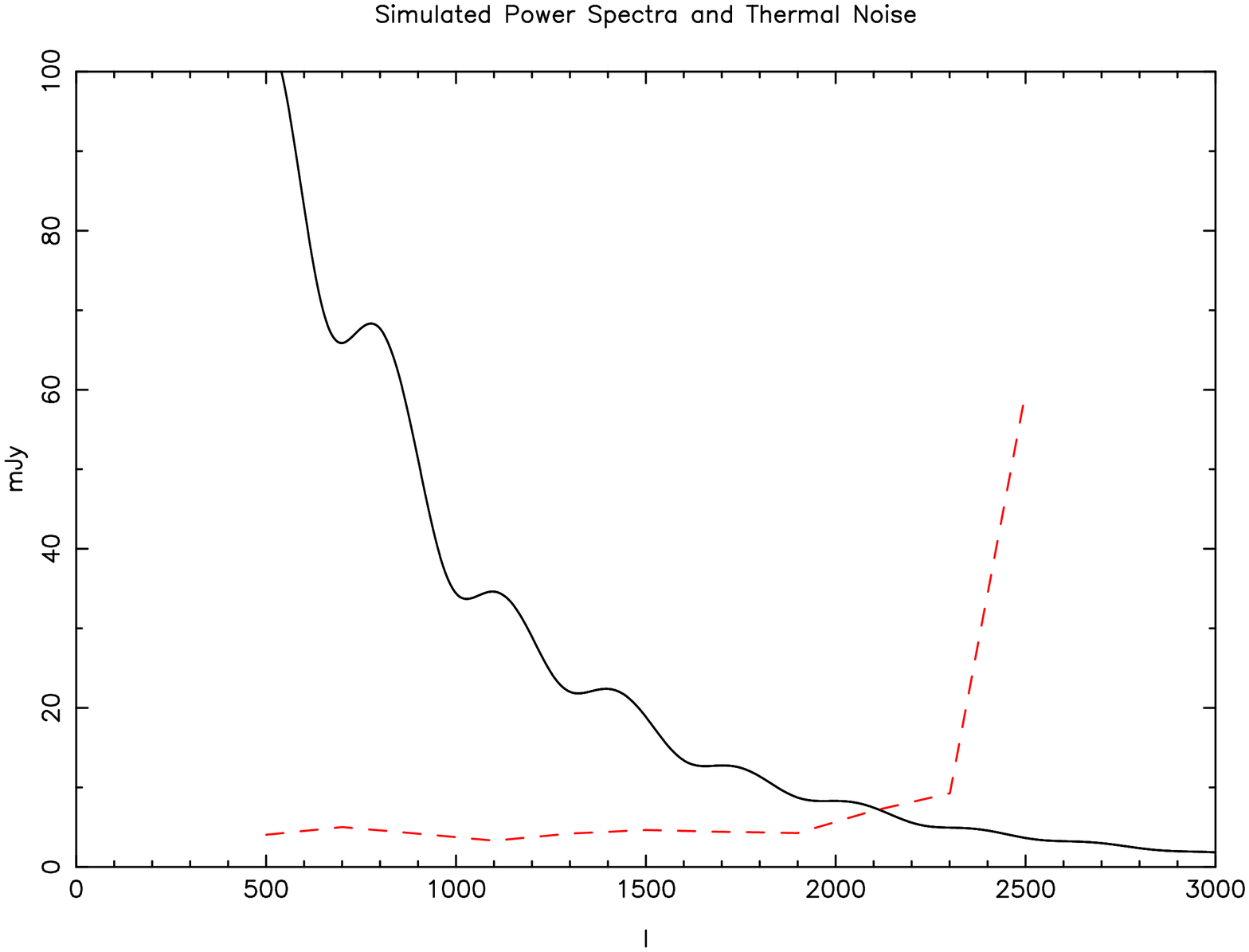,angle=-90,width=7cm,clip}}
\caption{Left: Proposed 'super-extended' array configuration with
  large 550-mm mirrors. Right: Expected flux sensitivity of the
  upgraded VSA in mJy versus $\ell$ (dashes). The
  expected rms signal from the CMB primary anisotropies is also shown
  (solid).\label{fig:super-arrayfig}}
\end{figure}

\begin{figure}
\begin{center}
\epsfig{file=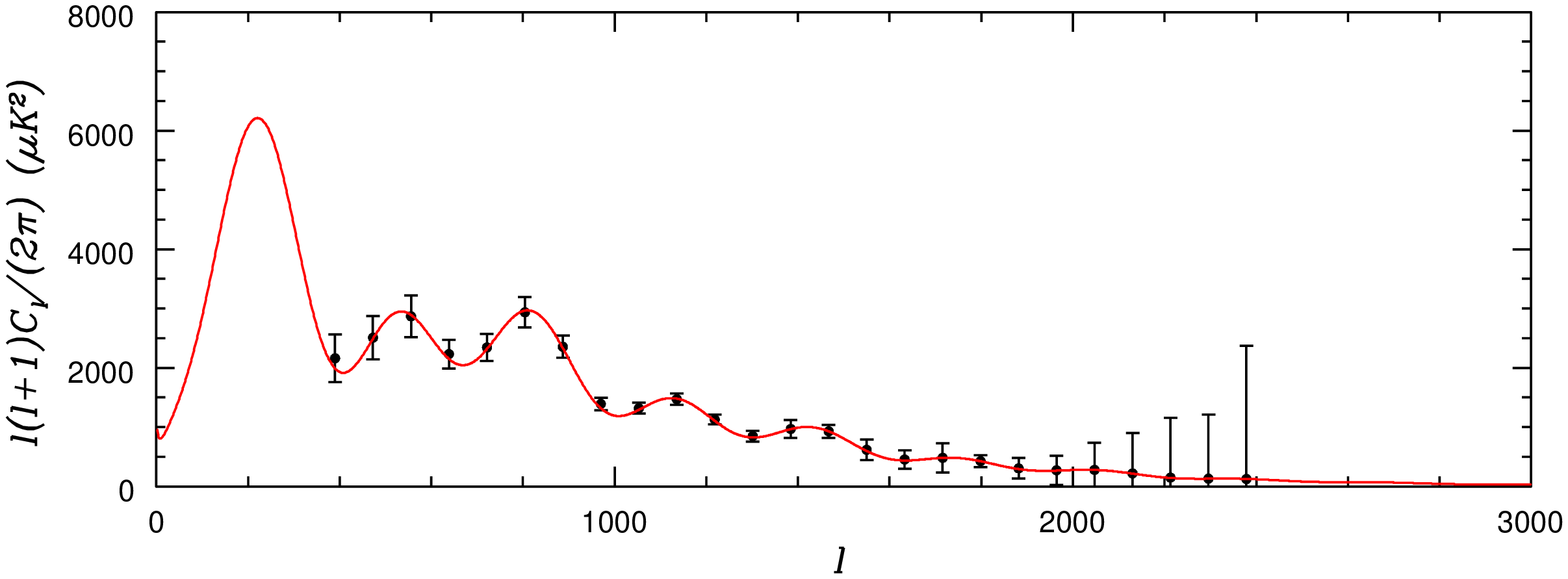,angle=-90,width=13cm}
\epsfig{file=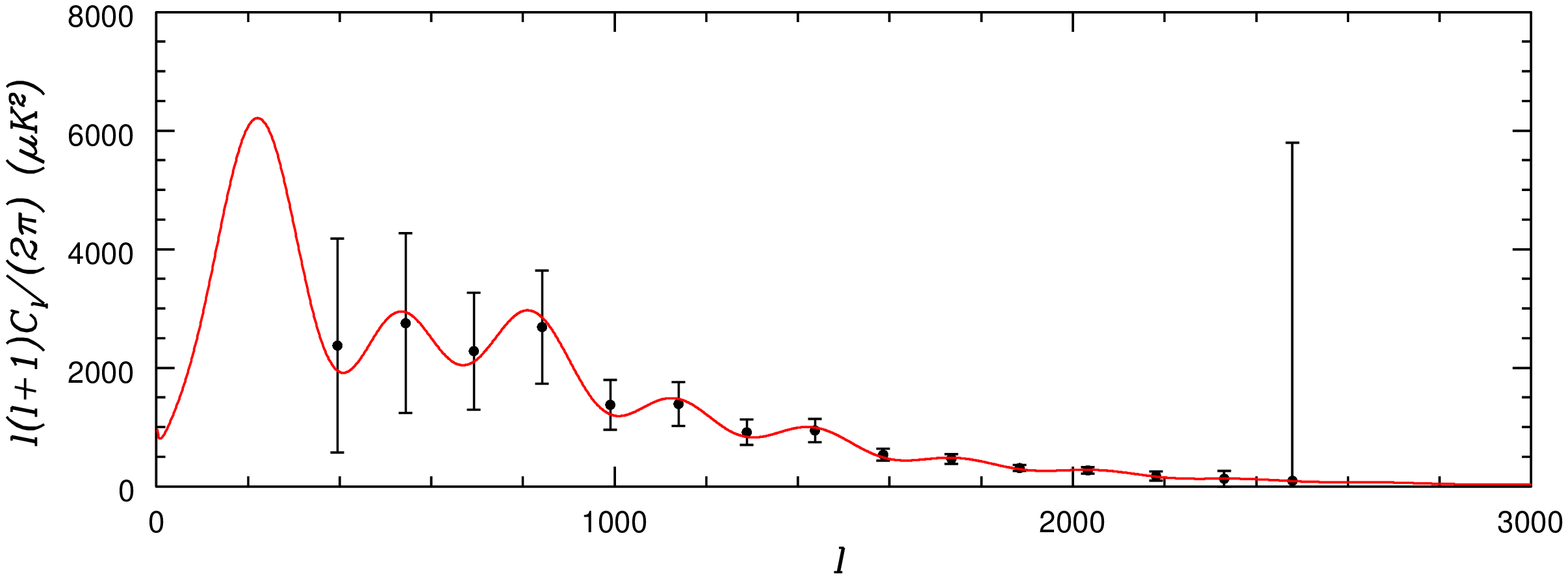,angle=-90,width=13cm}
\caption{Accuracy of CMB power spectrum recovery for the enhanced VSA 
in super-extended configuration after 12 months of observation. Top:
data from three 19-field mosaics. Bottom: data from a single deep
3-field mosaic.
\label{fig:sim_powerspectra}}
\end{center}
\end{figure}

We are now planning to upgrade the VSA to enable high resolution,
sensitive measurements of the power spectrum out to $\ell\sim2500$.
These measurements will be important both for breaking degeneracies in
cosmological model fitting and also for confirming or otherwise the
CBI measurement of excess power at high-$\ell$. To achieve the
required sensitivity and resolution, four upgrades to the current
system are proposed:

\begin{itemize}
\item{\textbf{Super-extended configuration} The current VSA array
configuration will be stretched to allow maximum baseline lengths of up to
3.0~m (see Figure \ref{fig:super-arrayfig}), and will allow observations out to
$\ell\sim2500$. To maintain a good filling factor and to increase the
sensitivity of the array, it is also necessary to replace the current 350-mm
antennas with larger light-weight antennas approximately 550-mm wide.  The new
antennas will be fabricated from carbon-fibre laminate with a metal mesh front
reflecting surface.}

\item{\textbf{Broadbanding the main array} At present the performance
    of the VSA is limited by its RF bandwidth. Using a broadband
    correlator and IF system already developed for AMI (see Section
    \ref{AMI}), we will upgrade the VSA from 1.5 to 6~GHz-bandwidth.
    This improvement will increase the observing speed of the VSA by a
    factor of 4.}

\item{\textbf{Improved system temperature} To further increase the sensitivity
and observing efficiency of the VSA, the front-end amplifiers of each receiver
will be upgraded by replacing the transistors with newer, lower-noise
devices. This will reduce the overall system temperature from the current 35~K
to 25~K. The combined improvement provided by the new broadband-system and
amplifiers will result in an 8-fold increase in the observing speed of the
instrument. The overall flux sensitivity of the upgraded VSA is shown
in Figure \ref{fig:super-arrayfig}.}

\item{\textbf{Improved source-subtraction sensitivity} Contamination of the
CMB power spectrum from extragalactic point sources increases as $\ell^2$ and
is expected to dominate at high $\ell$. Therefore the flux density down to
which point sources are currently detected and removed from VSA data must
be improved.  The single-baseline source-subtraction interferometer operating
alongside the VSA will therefore also be upgraded with a new broadband
correlator and improved amplifiers.}
\end{itemize} 
The new super-extended and enhanced VSA should be ready for operation by the
end of 2004. During the first 12 months of observation in the new
configuration, two distinct observing modes will be implemented. We plan
to observe three large regions of sky, each consisting of a 19-field hexagonal
close-packed mosaic.  This level of mosaicing will provide resolution in
$\ell$-space of $\Delta\ell=80$. In addition we will also make deeper
observations of a smaller patch of sky consisting of one 3-field mosaic. This
will be sample-variance limited at low $\ell$ but have sufficiently good
thermal noise to determine the CMB power level in broad bins ($\Delta \ell
\sim 250$) at $\ell > 2000$. Figure \ref{fig:sim_powerspectra} shows simulated power
spectra for these two observing modes. The error bars include sample variance,
thermal noise and calibration error.

%%%%%%%%%%%%%%%%%%%%%%%%%%%%%%%%%%%%%%%%%%%%%%%%%%%%%%%%%%%%%%%%%%%%%%%%%%%%%
\section{The Arcminute MicroKelvin Imager}
\label{AMI}
The Arcminute MicroKelvin Imager (AMI) is designed principally to
image secondary anisotropies in the CMB, at higher angular resolution
than the VSA. It consists of two interferometric arrays sited in
Cambridge, both operating in the frequency range 12--18~GHz and both
employing an analogue correlator system working over 6--12~GHz. The
smaller array is entirely new, consisting of ten 3.7-m paraboloidal
antennas situated inside a 30 $\times$ 40-m, 4.5-m high metal
enclosure.  The second array consists of the eight existing 13-m
diameter antennas of the Ryle Telescope reconfigured from the present
east-west arrangement into a new 2-dimensional compact array, with
their receivers upgraded, and the entire existing correlator system
replaced with the new 6--12~GHz design. This dual array design gives
AMI good temperature sensitivity over a very large range of angular
scales, as well as the high flux sensitivity needed to identify and
remove the effect of radio sources from the data. The novel correlator
design has all the intermediate frequency (IF) processing
(amplification, path compensation, slope equalisation, as well as the
correlation) done at the relatively high frequency of 6--12~GHz, using
microstrip circuits. The advantage of this is that the relatively low
fractional bandwidth eases the design of many of the system
components, compared with a more traditional baseband design. The
relatively low operating frequency allows a low system temperature
from a sea-level site, at the expense of increased confusion from
radio sources. This is dealt with by the longer baselines of the
larger array, which have high sensitivity to point sources, allowing
their removal from the short baseline data.

\begin{figure}[t]
\centerline{
\psfig{file=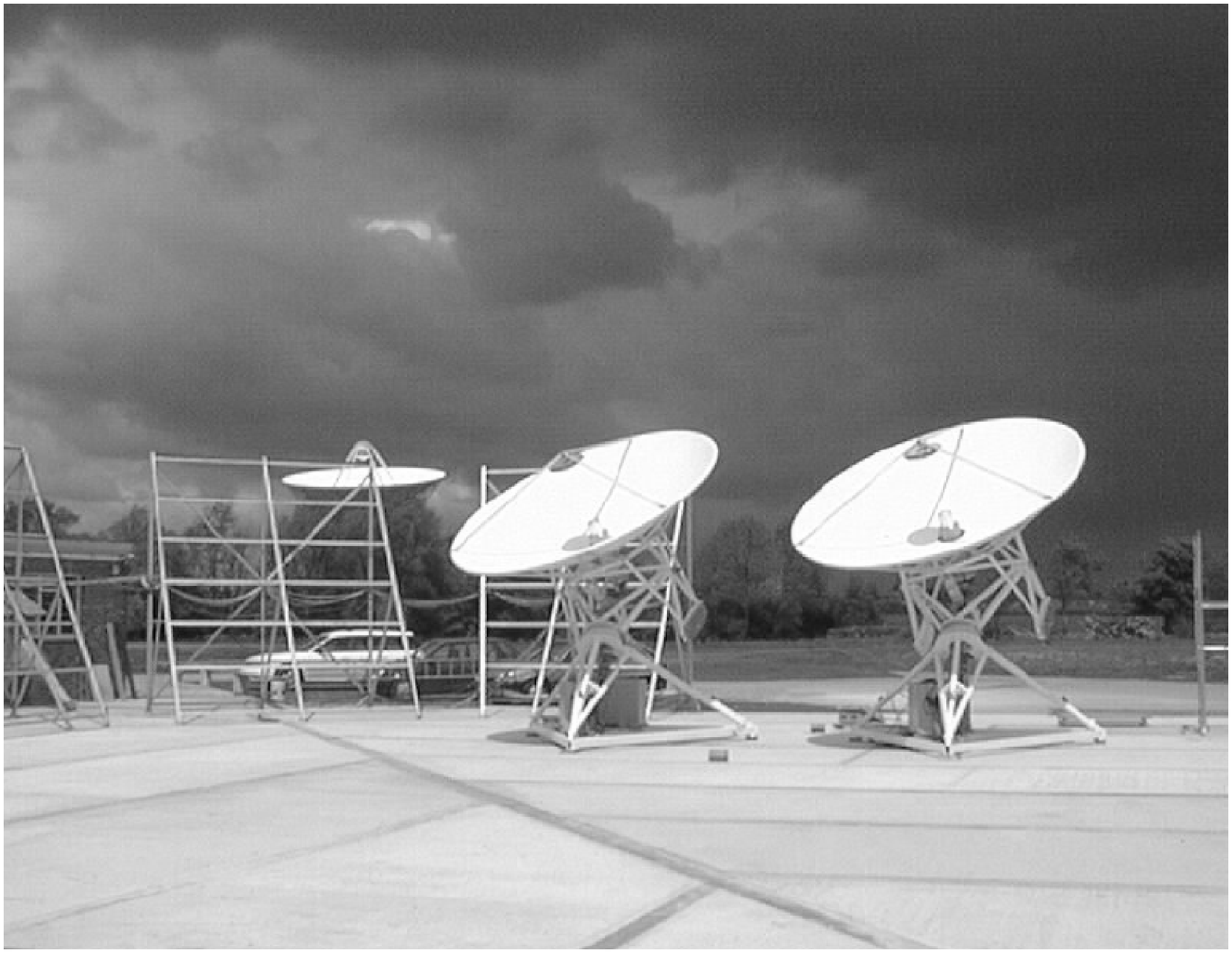,angle=-90,width=6cm}
\quad
\psfig{file=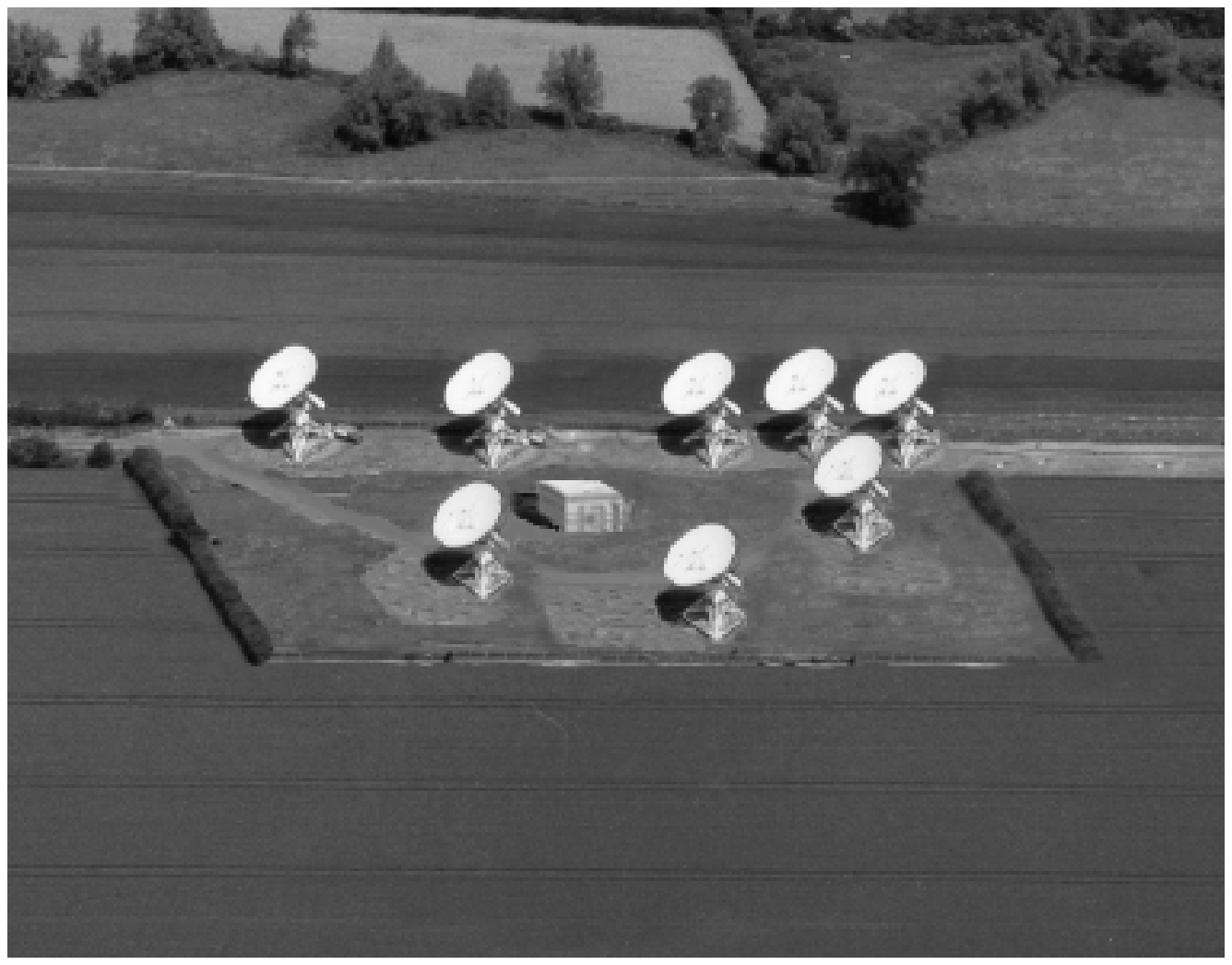,angle=-90,width=6cm}}
\caption{Left: Two of the ten 3.7-m antennas. Right: Artists impression of
the final configuration of the 13-m antennas. \label{ryle}}
\end{figure}

\begin{figure}[t]
\begin{center}
\epsfig{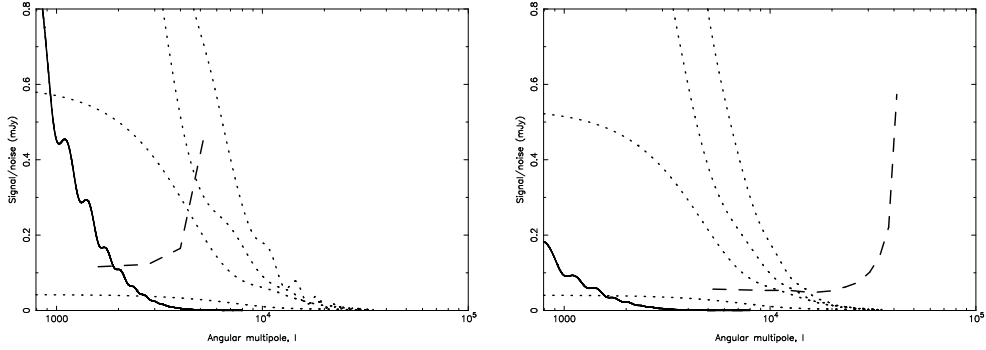}
\caption{The flux sensitivity of both the AMI 10-element, 3.7-m array (left)
and the 8-element, 13-m array (right), plotted as a function of angular
multipole, $\ell$. In each case the thermal sensitivity of each array
configuration (dashed line) has been calculated for a 6-month survey of a
$2^{\circ}\times2^{\circ}$ patch of sky. The predicted rms contribution from
the CMB (solid line) and the expected SZ signal from four clusters (dotted
lines) are also shown. The clusters are at redshifts $z = $0.35 -- 0.55 and
have masses of 5.2, 3.2, 1.7 and 0.33 $\times 10^{14} M_{\odot}$.
\label{fig:ryle_ami_sens} }
\end{center}
\end{figure}

The primary goal of AMI is to make blind surveys for galaxy clusters
via the SZ effect. Figure \ref{fig:ryle_ami_sens} shows the sensitivity of the two arrays plotted as flux density as a function of
$\ell$ (which is related to baseline length in wavelengths $u$ by
$\ell = 2 \pi u$). The thermal noise of the instrument, binned into
quasi-independent (aperture-sized) bins, is shown, along with the
expected rms signal from the primary CMB anisotropies, and the SZ
signal from a set of representative clusters. It can be seen that most
clusters are detected well above both the thermal noise and the
primary CMB signal in the 3.7-m array and on the shortest baselines of
the 13-m array, while the longer baselines of the 13-m array resolve
out both primary and secondary CMB anisotropies and provide
sensitivity to point sources (which have constant flux density as a
function of baseline).

AMI is currently under construction and is expected to begin operation towards
the end of 2003.

\begin{ack}
Many thanks to the large number of people working on the VSA and AMI
projects.  I would also like to thank Ben Rusholme for presenting this
talk on my behalf at very short notice.
\end{ack}

\end{document}